\documentclass[runningheads]{llncs}
\usepackage{hyperref}

\usepackage{color}

\usepackage{tikz}		
\usetikzlibrary{automata,shapes,arrows,positioning}

\usepackage{graphicx}
\usepackage{float}      
\usepackage{subfigure}

\usepackage{hyperref}	
\usepackage[british]{babel}

\usepackage{amsmath,amssymb}

\newcommand{\dist}{\mathit{Dist}}
\newcommand{\val}{\mathit{val}}

\title{Autonomous Agent Behaviour Modelled in PRISM -- A Case Study}
\titlerunning{Autonomous Agent Behaviour Modelled in PRISM}
\authorrunning{Hoffmann, Ireland, Miller, Norman and Veres}
\author{Ruth~Hoffmann\inst{1} \and Murray~Ireland\inst{1} \and Alice~Miller\inst{1} \and Gethin~Norman\inst{1} \and Sandor~Veres\inst{2}}
\institute{University of Glasgow, Glasgow, G12 8QQ, Scotland \and
	University of Sheffield, Sheffield, S1 3JD, England}

\begin{document}
\maketitle
\begin{abstract}
Formal verification of agents representing robot behaviour is a growing area due to the demand that  autonomous systems have to be proven safe. In this paper we present an abstract definition of autonomy which can be used to model autonomous scenarios and propose the use of small-scale simulation models representing abstract actions to infer quantitative data.
To demonstrate the applicability of the approach we build and verify a model of an unmanned aerial vehicle (UAV) in an exemplary autonomous scenario, utilising this approach.
\end{abstract}

\section{Introduction}
Autonomous systems have the ability to decide at run-time what to do
and how to do it. A critical question is how this decision making
process is implemented.

Increasingly, autonomous systems are being deployed
within the public domain (e.g. driverless cars, delivery
drones). Naturally, there is concern that these systems are reliable,
efficient and - most of all - safe. Although testing is a necessary
part of this process, simulation and formal verification are key
tools, especially at the early stages of design where experimental testing is both
infeasible and dangerous. Simulation allows us to view the continuous dynamics and monitor behaviour of a system. On the other hand, model checking allows us to formally verify properties of a finite representation. Whereas the simulation model is close to an implementation, simulation runs are necessarily incomplete. Verification models, on the other hand, require us to abstract more coarsely.

The decisions made by an autonomous agent depend on the current state of the environment, specifically in terms of data perceived by the agent from its sensors. If model checking is to be used for the verification of autonomous systems we must reflect the uncertainty associated with the state of the environment by using probabilistic model checking.

We propose a framework for analysing autonomous systems, specifically to investigate decision-making, using probabilistic model checking of an abstract model where quantitative data for abstract actions is derived from small-scale simulation models. We illustrate
our approach for an example system composed of a UAV searching for and
collecting objects in an arena. The simulation models for abstract actions are generated using the object-oriented framework Simulink and the abstract models are specified and verified using the probabilistic model checker PRISM. In our example, autonomous
decision making involves making a weighted choice between a set of
possible actions, and is loosely based on the Belief-Desire-Intention
architecture~\cite{VMLM11}.

Previous work in which model checking is used to verify autonomy includes \cite{Dennis2014} in which the decision making process is verified in isolation, while our aim is to integrate this process with the autonomous agent as a whole. Other research includes an investigation of the cooperative behaviour of robots, where each robot is represented by a hybrid automaton~\cite{CCK03}, and verification of a consensus algorithm using experimental verification and an external observer~\cite{CH10}.

\section{Autonomy}
\label{sec:formalaut}
In order to formally define autonomous behaviour, we introduce finite state machines which abstract the autonomous actions independent of the agent type.

Before we give the formal definitions we require the following notation. For a finite set of variables $V$, a valuation of $V$ is a function $s$ mapping each variable in $V$ to a value in its finite domain. Let $\val(V)$ be the set of valuations of $V$. For any $s \in \val(V)$, $v \in V$ and value $x$ of $V$, let $s[v{:=}x]$ and $s[v{\pm}x]$ be the valuations where for any $v' \in V$ we have $s[v{:=}x](v')=x$ and $s[v{\pm}x](v')=s(v'){\pm}x$ if $v'{=}v$ and $s[v{:=}x](v')=s[v{\pm}x](v')=s(v')$ otherwise. For a finite set $X$, a probability distribution over $X$ is a function $\mu : X \rightarrow [0, 1]$ such that $\sum_{x \in X} \mu(x) = 1$. Let $\dist(X)$ be the set of distributions over $X$.

\begin{definition}
	A \emph{probabilistic finite-state machine} is a tuple $\mathcal{M} {=} (V,I,A,T)$ where:
	$V$ is a finite set of \emph{variables}; $I\subseteq \val(V)$ a set of \emph{initial states}; $A$ a finite set of \emph{actions} and $T: \val(V) {\times} A \rightarrow \dist(S)$ a (partial) \emph{transition function}.
\end{definition}
The set of states of $\mathcal{M} {=} (V,I,A,T)$, denoted $S$, is the set of valuations $\val(V)$ of $V$. Let $A(s)$ denote the actions available from state $s$, i.e. the actions $a \in A$ for which $T(s,a)$ is defined. In state $s$ an action is chosen non-deterministically from the available actions $A(s)$ and, if action $a$ is chosen, the transition to the next state is made according to the probability distribution $T(s,a)$. A probabilistic finite-state machine describes a system without autonomy, we introduce this through a weight function adding decision making to the finite-state machine.

\begin{definition}\label{afsm-def}
An \emph{autonomous} probabilistic finite-state machine is a tuple $\mathcal{A} = (V,I,A,T,w)$ where $(V,I,A,T)$ is a probabilistic finite-state machine and $w$ is a \emph{weight function} $w: \val(V) {\times} A \rightarrow [0,1]$ such that for any $s \in \val(V)$ and $a \neq b \in A$ we have $w(s,a) \neq w(s,b)$ and $w(s,a){>}0$ implies $a \in A(s)$.
\end{definition}
In an autonomous machine the non-determinism in the first step of a transition is removed. More precisely, if a machine $\mathcal{A} {=} (V,I,A,T,w)$ is in state $s$, then the action performed is that with the largest weight, that is the action:
\[ \begin{array}{c}
a_{s,w} = \mathrm{arg}\,\max \{ w(s,a) \mid a \in A(s) \} \, .
\end{array} \]
Requiring the weights for distinct actions and the same state to be different ensures this action is always well defined. Having removed the non-determinism through the introduction of a weight function, the semantics of an autonomous finite state machine is a discrete time Markov chain.

\section{UAV Example}
\label{sec:model}
In this case study we consider a specific example (a simple search and retrieve example, with a UAV in a finite sized arena) to demonstrate our approach.

\begin{figure}[!t]
\centering
\begin{tikzpicture}[auto,node distance=1.5cm,on grid,>=latex,scale=0.3, font=\scriptsize,every state/.style={rectangle}, every node/.style={align=center}]

\node[initial,state,initial where=above] (S0) {Idle \\ 0};
\node[state] (S1) [right =2.5cm of S0] {Take-off \\ 1};
\node[state] (S2) [below of=S1] {Initialise \\ 2};
\node[state] (S14) [below of=S2,text width=1.3cm] {Go to Search \\ 14};
\node[state] (S3) [below of=S14] {Search \\ 3};
\node[state] (S4) [below of=S3,text width=1.5cm] {Target Approach \\ 4};
\node[state] (S5) [below of=S4,text width=1.4cm] {Descend to grab \\ 5};
\node[state] (S6) [below of=S5] {Grab \\ 6};

\node[state] (S13) [left=2.5cm of S0] {Land \\ 13};
\node[state] (S12) [below of=S13] {Return \\ 12};
\node[state] (S11) [below of=S12] {Ascend \\ 11};
\node[state] (S10) [below of=S11] {Deposit \\ 10};
\node[state] (S9) [below of=S10,text width=1.5cm] {Descend to deposit \\ 9};
\node[state] (S8) [below of=S9] {Transport \\ 8};
\node[state] (S7) [below of=S8] {Ascend \\ 7};

\node[state] (S15) [left=2.5cm of S5,text width=1.5cm] {Return to Object \\ 15};

\node[state,ellipse] (All1) [left = 2.5cm of S12] {$\{1,\ldots,11,$\\$14,15\}$};

\node[state] (S16) [right=2.5cm of S14] {Emergency \\ Land \\ 16};
\node[state,ellipse] (All2) [above of=S16] {$\{0,\ldots,15\}$};

\path[->]
(S0) edge (S1)
(S1) edge (S2)
(S2) edge (S13)
	edge (S14)
(S3) edge [loop right] (S3)
	edge (S4)
	edge (S12)
(S4) edge (S5)
(S5) edge (S6)
(S6) edge (S7)
(S7) edge (S8)
	edge (S15)
(S8) edge [loop left] (S8)
    edge (S9)
	edge (S15)
	edge [bend left,out=50,in=130] (S12)
(S9) edge (S10)
(S10) edge (S11)
(S11) edge (S12)
    edge (S14)
(S12) edge (S13)
(S13) edge (S0)
(S14) edge (S3)
(S15) edge (S4)
    edge (S14)
(All1) edge (S12)
(All2) edge (S16);

\end{tikzpicture}
\vspace*{-0.5cm}
\caption{The finite state machine representing the scenario.}\label{fig:FSM}
\vspace*{-0.4cm}
\end{figure}
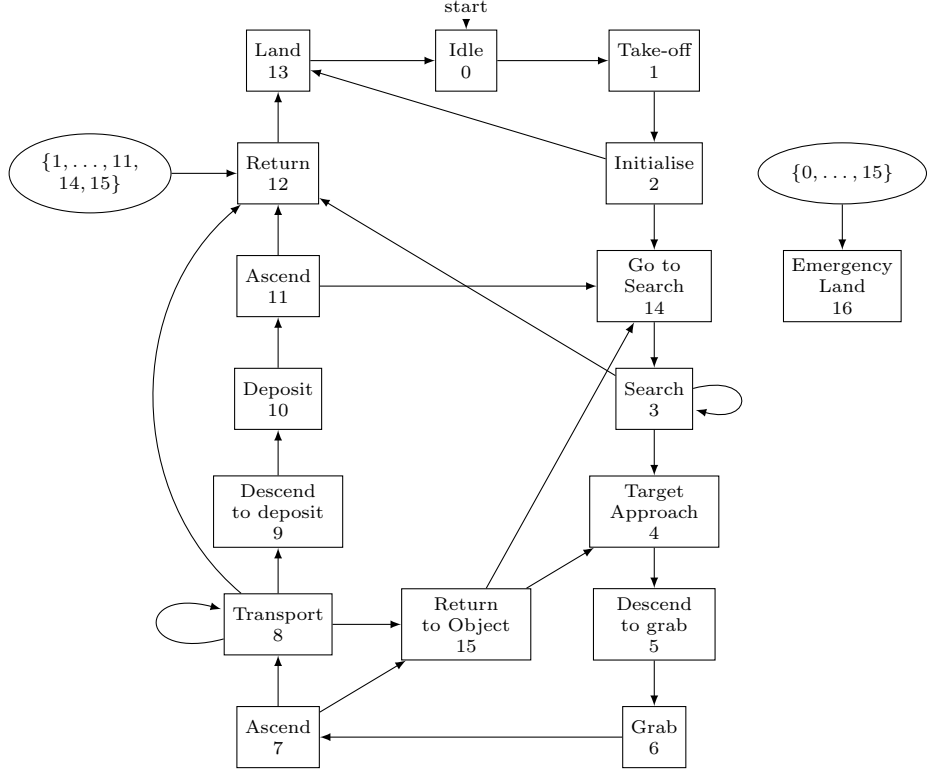

The UAV first takes off and checks whether the system and sensors are functional. If the UAV detects an issue in the system, then it returns to base. Otherwise it will proceed to search for a given number of objects. When an object is found, the UAV positions itself above the object and descends until the grabber can pick it up. The UAV then ascends to transportation height and transports the object to the deposit site. There is the possibility that the UAV will drop its object along the way and need to retrieve it. Once the UAV is above the deposit site, it releases the object and ascends back to search height. It will then decide whether it continues the search or returns to the base and complete the mission. During operation, the UAV may return to base to recharge if it is low on battery, or conduct an emergency landing, due to an internal system error. If the mission time limit  is reached, the UAV abandons the mission and returns to base. Figure~\ref{fig:FSM} represents this scenario, showing the different modes of the UAV and progression between the modes.

We represent this scenario using a autonomous finite-state machine $\mathcal{A}$. The variables $V$ of $\mathcal{A}$ are given by:
\begin{itemize}
\item $\mathit{obj}$ the number of objects which have not been found;
\item $\mathit{pos} {=} (\mathit{pos}_x,\mathit{pos}_y)$ the position of the UAV in the arena;
\item $\mathit{ret} {=} (\mathit{ret}_x,\mathit{ret}_y)$ the return coordinates when search is interrupted;
\item $m$ the current mode of the UAV;
\item $t$ the mission time;
\item $b$ the battery charge level.
\end{itemize}
Each state $s\in val(V)$ of $\mathcal{A}$ is a valuation of these variables. The transition and weight functions $T$ and $w$ are based on Figure~\ref{fig:FSM}. We focus on the target approach and search modes of the UAV.

In the target approach mode ($m{=}4$), the UAV positions itself above an observed object and we denote this abstract action by $\mathit{Approach}$. Thus, the weight function is $w(s,\mathit{Approach}){=}1$ for all states $s$ such that $s(m){=}2$, and for any such state $s$ we have for any $s' \in S$:
\[\begin{array}{c}
T(s,\mathit{Approach})(s') =
\begin{cases}
1 & \mbox{if} \; s' = s[m{:=}5][t{+}T_\mathit{ap}][b{-}B_\mathit{ap}] \\
0 & \mbox{otherwise}
\end{cases}
\end{array}
\]
where $T_\mathit{ap}$ and $B_\mathit{ap}$ are the time and battery charge used approaching the object, and $m{=}5$ is the mode for descending. The abstract action $\mathit{Approach}$ models several different operations of the UAV including the use of its camera and navigational system. A small-scale simulation model was built for this abstract action to provide the required quantitative data.

When the UAV is in search mode ($m{=}3$) there are two actions that can occur: $\mathit{Search}$ and $\mathit{BatteryLow}$ with the UAV continuing search if the battery charge level is above a certain threshold, and returning to base otherwise. The weight function for the $\mathit{Search}$ action from any state $s$ such that $s(m){=}3$ is given by:
\[ \begin{array}{c}
w(s,\mathit{Search})=
\begin{cases}
0 & \text{if } s(b) \leq B_\mathit{low} \\
1 & \text{if } s(b) > B_\mathit{low}
\end{cases}
\end{array}
\]
and for the $\mathit{BatteryLow}$ action we have $w(s,\mathit{BatteryLow}) = 1{-}w(s,\mathit{Search})$.
Concerning the transition function we have for any $s' \in S$:
\[\begin{array}{c}
T(s,\mathit{Search})(s') =
\begin{cases}
 1-\alpha & \mbox{if} \; s' =s [\mathit{pos} {:=} \Delta \mathit{pos}][t{+}\Delta t][b{-}\Delta b] \\
\alpha & \mbox{if} \; s' =s [\mathit{pos} {:=} \Delta \mathit{pos}][\mathit{ret} {:=} \Delta \mathit{pos}][m{:=}4][t{+}\Delta t][b{-}\Delta b] \\
0 & \mbox{otherwise}
\end{cases}
\end{array}
\]
and $T(s,\mathit{BatteryLow})(s){=} 1$ if $s' {=}s [\mathit{pos} {:=} \Delta \mathit{pos}][\mathit{ret} {:=} \Delta \mathit{pos}][m{:=}12][t+{}\Delta t][b{-}\Delta b]$ and 0 otherwise,
where $\Delta pos$ denotes the movement from one discrete square in the arena to the next, $\Delta t$ and $\Delta b$ are the time and battery consumption of the UAV while moving one square. The UAV has probability $\alpha$ of finding an object in a given position, if an object is found the UAV changes to mode $m{=}4$ and $\mathit{ret}$ is set to the current coordinates, as the search has been interrupted. If no object is found, the UAV continues searching.

\section{Results}
\label{sec:res}

We have modelled our scenario in the probabilistic model checker PRISM~\cite{KNP11}, building small-scale simulation models to determine individual abstract actions to generate probabilistic and timing values. To encode the timing values in PRISM as integer variables we take the floor and ceiling, introducing non-determinism into the model and upper and lower bounds for properties~\cite{KKNP10}. The experiments were performed on a computer with 16GB RAM and a 2.3 GHz Intel Core i5 processor.

The main properties of interest concern the UAV successfully completing the mission (finding and depositing all objects within the time limit) and failing the mission (either due to an emergency landing, missing some objects or running out of time). We have also considered other properties including how often the UAV drops an object and how often it recharges during a mission, more details can be found in the repository~\cite{Repo}.

We have analysed two scenarios where there are 3 objects and a time limit of 900s, and 2 objects and a time limit of 500s respectively. For the first scenario the model has $116\,191\,709$ states, was built in $488$s and verifying a property varied between $298$s and $813$s. This is far more efficient than running Monte Carlo simulations, as simulating $10\,000$ runs of the same scenario takes over two weeks. For the second scenario the model has $35\,649\,503$ states and model construction time was $77$s.

\begin{figure}[t]
    \centering
    {{\subfigure[Scenario 1.]{\includegraphics[scale=0.39]{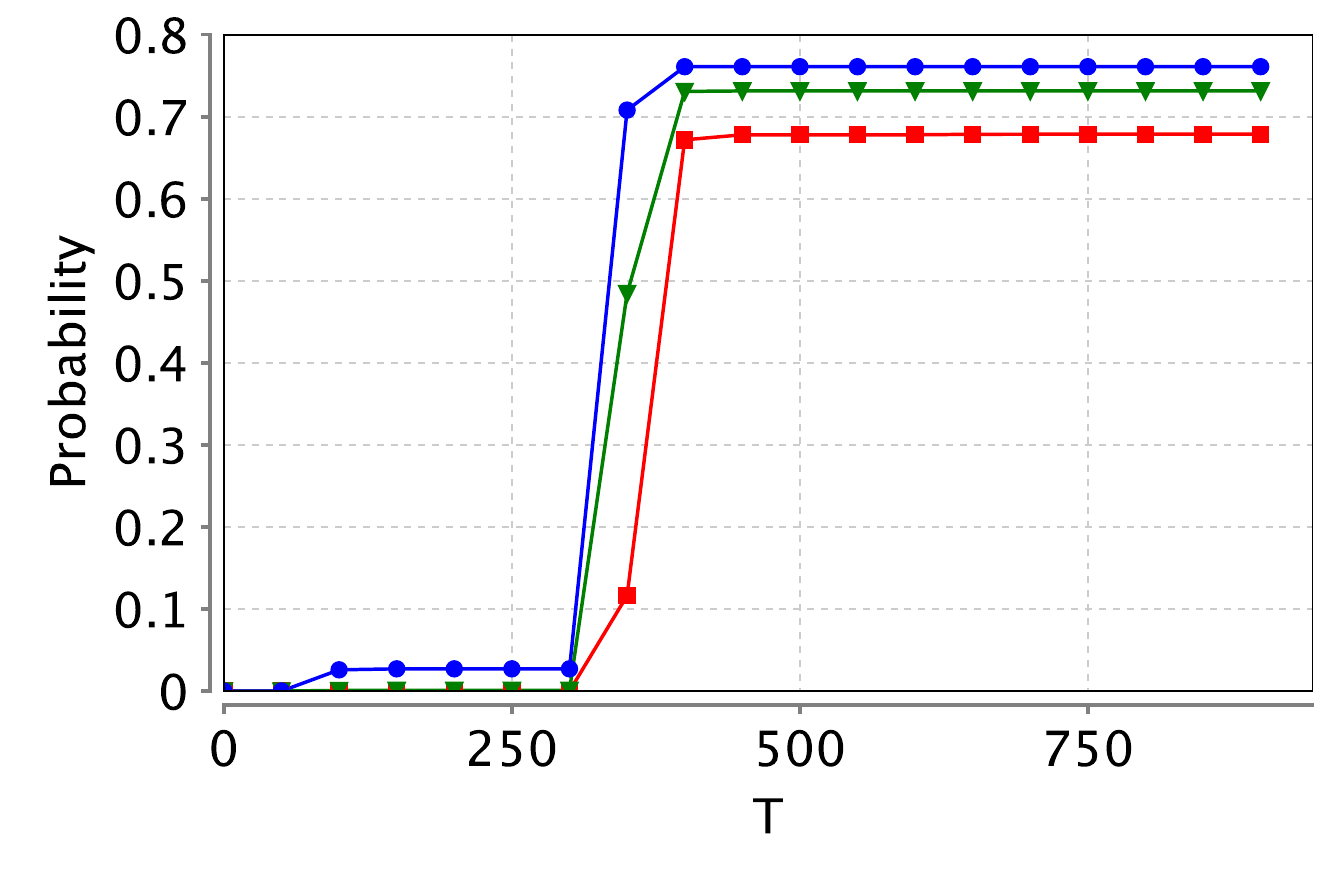}}}
    {\subfigure[Scenario 2.]{\hspace*{-0.2cm}\includegraphics[scale=0.39]{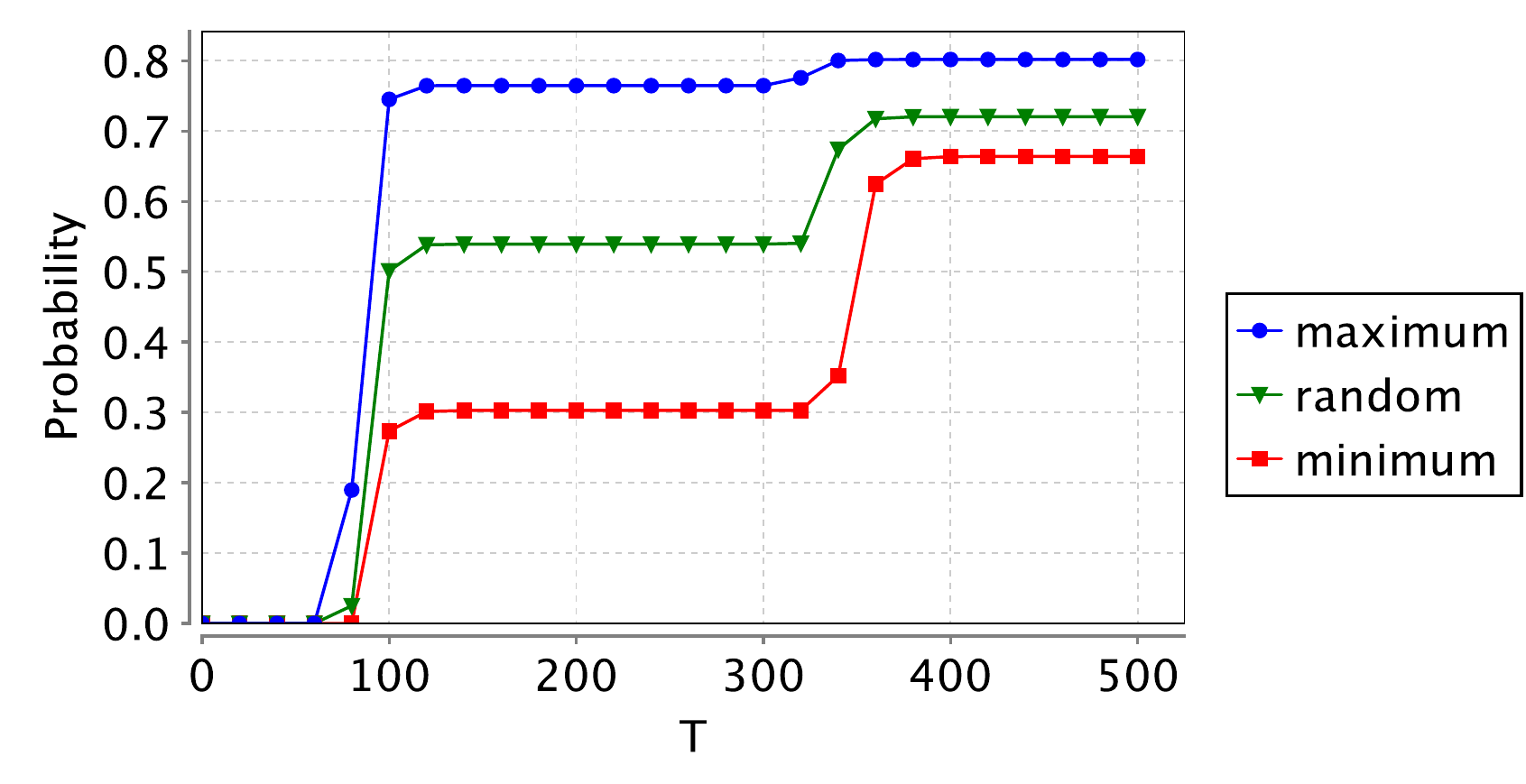}}}}
    \vspace*{-0.4cm}
    \caption{Probability of completing the mission successfully by deadline $T$.}\label{fig:deadline}
    \vspace*{-0.4cm}
\end{figure}

For the first scenario, the maximum and minimum probabilities of the UAV completing a mission are $0.7610$ and $0.6787$ respectively. The maximum and minimum probabilities of running out of time are negligible, searching the arena and missing some objects are $0.2617$ and $0.1808$, and $0.0628$ and $0.0552$ for performing an emergency landing. Figure~\ref{fig:deadline} shows the maximum and minimum probability of a successful mission within a time bound as well as the results obtained when the non-determinism is replaced by a uniform random choice. The probability increases after a threshold time as the UAV has to search a proportion of the arena before finding all objects.

\section{Conclusions}
\label{sec:concl}

We have proposed using small-scale simulation models to inform probabilistic models used for verification. The simulation models can be used to provide quantitative data for abstract actions. The probabilistic models can be used for fast property specific verification, that is not possible using simulations alone.

Our approach is highly adaptable; once the initial small-scale simulation and probabilistic models have been set up, different decision algorithms can be easily substituted and analysed. Our example illustrates the use of a weight function for decision making. In a more extensive scenario the weight function would be more complex (e.g.\ involving current values associated with all sensors and guiding systems). Our use of non-determinism when approximating quantitative data obtained from the small-scale simulation models allows us to provide an range of uncertainty for our results. We aim to formally prove a link between the simulation and the abstract model to allow us to infer results from the abstract model for the actual system. To allow the analysis of more complex scenarios we plan to incorporate abstraction techniques.
\vskip6pt \noindent
{\bf Acknowledgments.}
This work was supported by the Engineering and Physical Sciences Research Council [grant number EP/N508792/1].

\bibliographystyle{splncs03}
\bibliography{library}

\end{document}